\documentstyle[prd,aps,preprint,psfig]{revtex}

\newcommand{\CL}{{\cal L}}

\newcommand{\CO}{{\cal O}}

\newcommand{\bear}{\begin{array}}  \newcommand{\eear}{\end{array}}
\newcommand{\bea}{\begin{eqnarray}}  \newcommand{\eea}{\end{eqnarray}}
\newcommand{\beq}{\begin{equation}}  \newcommand{\eeq}{\end{equation}}
\newcommand{\bef}{\begin{figure}}  \newcommand{\eef}{\end{figure}}
\newcommand{\bec}{\begin{center}}  \newcommand{\eec}{\end{center}}
\newcommand{\non}{\nonumber}  
\newcommand{\lmk}{\left(}  \newcommand{\rmk}{\right)}
\newcommand{\lkk}{\left[}  \newcommand{\rkk}{\right]}
\newcommand{\lhk}{\left \{ }  \newcommand{\rhk}{\right \} }
\newcommand{\del}{\partial}  

\newcommand{\bib}{\bibitem} 
\newcommand{\la}{\left\langle} \newcommand{\ra}{\right\rangle}

\def\IB#1#2#3{{\bf #1}, #2 (19#3)}
\def\IBID#1#2#3{{\it ibid}. {\bf #1}, #2 (19#3)}

\def\APJ#1#2#3{Astrophys. J. {\bf #1}, #2 (19#3)}

\def\CQG#1#2#3{Class. Quantum Grav. {\bf #1}, #2 (19#3)}
\def\CQGG#1#2#3{Class. Quantum Grav. {\bf #1}, #2 (20#3)}

\def\JHEP#1#2#3{J. High Energy Phys. {\bf #1}, #2 (19#3)}
\def\JL#1#2#3{JETP. Lett. {\bf #1}, #2 (19#3)}

\def\MPLA#1#2#3{Mod. Phys. Lett. A {\bf #1}, #2 (19#3)}

\def\NPB#1#2#3{Nucl. Phys. {\bf B#1}, #2 (19#3)}

\def\PLB#1#2#3{Phys. Lett. B {\bf #1}, #2 (19#3)}

\def\PLBold#1#2#3{Phys. Lett. {\bf#1B}, #2 (19#3)}

\def\PRD#1#2#3{Phys. Rev. D {\bf #1}, #2 (19#3)}
\def\PRDD#1#2#3{Phys. Rev. D {\bf #1}, #2 (20#3)}

\def\PRL#1#2#3{Phys. Rev. Lett. {\bf#1}, #2 (19#3)}
\def\PRLL#1#2#3{Phys. Rev. Lett. {\bf#1}, #2 (20#3)}

\def\PRT#1#2#3{Phys. Rep. {\bf#1}, #2 (19#3)}
\def\PTP#1#2#3{Prog. Theor. Phys. {\bf #1}, #2 (19#3)}
\def\PZE#1#2#3{Pis'ma Zh. \'Eksp. Teor. Fiz. {\bf#1}, #2 (19#3)}

\begin{document}

%@twocolumnfalse\endcsname
%%
%%
\tighten
\draft
\title{Natural Chaotic Inflation in Supergravity and Leptogenesis}
\author{M. Kawasaki}
\address{Research Center for the Early Universe, University of Tokyo,
  Tokyo 113-0033, Japan}
\author{Masahide Yamaguchi}
\address{Research Center for the Early Universe, University of Tokyo,
  Tokyo 113-0033, Japan}
\author{T. Yanagida}
\address{Department of Physics, University of Tokyo, Tokyo 113-0033,
Japan \\ and \\ Research Center for the Early Universe, University of
Tokyo, Tokyo 113-0033 Japan}
%%%

\date{\today}

\maketitle

\begin{abstract}
    We comprehensively investigate a chaotic inflation model proposed
    recently in the framework of supergravity. In this model, the form
    of K\"ahler potential is determined by a symmetry, that is, the
    Nambu-Goldstone-like shift symmetry, which guarantees the absence
    of the exponential factor in the potential for the inflaton field.
    Though we need the introduction of small parameters, the smallness
    of the parameters is justified also by symmetries. That is, the
    zero limit of the small parameters recovers symmetries, which is
    natural in the 't Hooft's sense. The leptogenesis scenario via the
    inflaton decay in this chaotic inflation model is also discussed.
    We find that the lepton asymmetry enough to explain the present
    baryon number density is produced for low reheating temperatures
    avoiding the overproduction of gravitinos.
\end{abstract}


%]

%%%%%%%%%%%%%%%%%%%%%%%%%%%%%%%%%%%%%%%%%%%%%%%%%%%%%%%%%%%%%%%%%%%%%%
\section{Introduction}
%%%%%%%%%%%%%%%%%%%%%%%%%%%%%%%%%%%%%%%%%%%%%%%%%%%%%%%%%%%%%%%%%%%%%%

\label{sec:int}

Big-bang cosmology is a very attractive theory because it explains
well the three main observational results in cosmology, that is,
Hubble expansion, the cosmic microwave background radiation (CMBR),
and the primordial abundance of light elements. But it has famous
problems, namely, the horizon problem and the flatness problem, and
does not account for the origin of primordial fluctuations of CMBR as
observed by the Comic Background Explorer (COBE) satellite
\cite{COBE}.  The most natural solution to these problems is inflation
\cite{inflation}.  Until now, many types of inflation models have been
proposed. Among them, chaotic inflation is special in that it can take
place at about the Planck time. Other types of inflation occur
generally at much later times so that they suffer from the flatness
(longevity) problem\cite{inflation} though it is milder than the
original one, that is, why the universe lives so long up to the low
energy scale.  Furthermore, other types of inflation except chaotic
and topological inflation also suffer from the initial value problem
\cite{inflation,hi}, that is, why the inflaton field $\varphi$ is
homogeneous over the horizon scale and lies in the small region of the
potential which leads to a successful inflation. If the universe is
open at the beginning \cite{open}, the flatness problem may be evaded
and the topological inflation may occur.\footnote{Exactly speaking,
for a successful topological inflation in supergravity, the K\"ahler
potential must be fine-tuned against quantum corrections in order to
keep the flatness of the potential near the origin.} However, the
chaotic inflation gives the most natural solution to the above
problems since it takes place at about the Planck time. Thus, the
chaotic inflation is the most attractive inflation without any fine
tuning.

The fact that inflation takes place at higher energy scales than the
electroweak scale confronts us with a hierarchy problem between such
two energy scales. One of the most attractive solutions is
supersymmetry (SUSY) \cite{SUSY}, which stabilizes such a large
hierarchy against radiative corrections. Thus, it is important to
consider inflation in the framework of the local version of SUSY,
i.e., supergravity.

Chaotic inflation can be realized for a very simple polynomial
potential. Due to this simplicity, a lot of applications have been
investigated for the chaotic inflation, for example, preheating
\cite{preheating}, superheavy particle production \cite{X}, and
primordial gravitational waves \cite{Starobinsky}. It is, however,
very difficult to realize such a polynomial potential in supergravity
because the minimal supergravity potential has an exponential factor
($e^{\varphi^{\dagger}\varphi/M_{G}^{2}+\cdots}$), which prevents
inflaton $\varphi$ from having an initial value much larger than the
gravitational scale $M_{G} \simeq 2.4 \times 10^{18}$~GeV.  Thus, it
has been believed to be very difficult in incorporating the chaotic
inflation in the framework of supergravity. Although some models for
the chaotic inflation were proposed using specific K\"ahler potentials
instead of the canonical K\"ahler potential \cite{GL,MSYY}, such
K\"ahler potentials have no symmetry reason and we must invoke a fine
tuning.

However, we have recently constructed a natural chaotic inflation
model in supergravity without any fine tuning \cite{KYY}. The term
``natural'' has two meanings. First of all, the form of K\"ahler
potential is determined by a symmetry, that is, the
Nambu-Goldstone-like shift symmetry, which guarantees the absence of
the exponential factor in the potential for the inflaton field. Though
we need the introduction of small breaking parameters, the smallness
of parameters is justified also by symmetries. That is, the zero limit
of small parameters recovers symmetries, which is natural in the 't
Hooft's sense \cite{tHooft}. This is the second meaning of our term
``natural.'' In this paper, we comprehensively investigate this
chaotic inflation model, particularly paying attention to the small
parameters of symmetry breaking in the superpotential.

As an application of the above new type of chaotic inflation model
\cite{KYY}, we discuss the leptogenesis. Recent experimental results
on the atmospheric neutrinos strongly indicate that neutrinos have
small masses of the order of 0.01$-$0.1 eV \cite{SK}. Such small
masses are naturally explained by the seesaw mechanism \cite{seesaw},
which predicts superheavy right-handed neutrinos. The presence of
Majorana masses of right-handed neutrinos naturally leads to the
leptogenesis because it violates the lepton number conservation. The
decay of superheavy Majorana neutrinos produces the lepton number
asymmetry, in particular, $B-L$ asymmetry if $C$ and $CP$ symmetries
are broken, which is converted into baryon asymmetry \cite{FY} through
the sphaleron effects \cite{KRS}. Therefore, we discuss a leptogenesis
scenario in the above mentioned chaotic inflation model.

In the next section, we briefly review on the chaotic inflation model
in supergravity. In Sec. \ref{sec:dynamics}, we investigate the
dynamics of chaotic inflation. In Sec. \ref{sec:lepto}, we discuss the
leptogenesis via the inflaton decay. The last section is devoted to
discussion and conclusions.

%%%%%%%%%%%%%%%%%%%%%%%%%%%%%%%%%%%%%%%%%%%%%%%%%%%%%%%%%%%%%%%%%%%%%%
\section{Natural chaotic inflation model in supergravity}
%%%%%%%%%%%%%%%%%%%%%%%%%%%%%%%%%%%%%%%%%%%%%%%%%%%%%%%%%%%%%%%%%%%%%%

\label{sec:chao}

As explained in the introduction, the chaotic inflation is special in
that it takes place around the gravitational scale and hence it does
not suffer from the flatness (longevity) and the initial value
problems. But it was a long-standing problem to realize a chaotic
inflation naturally in supergravity because the minimal supergravity
potential has an exponential growth
($e^{\varphi^{\dagger}\varphi/M_{G}^{2}+\cdots}$) for the inflaton
field $\varphi$, which prevents the inflaton $\varphi$ from taking an
initial value much larger than the gravitational scale. However, we
have recently proposed a natural chaotic inflation model in
supergravity by imposing Nambu-Goldstone-like shift symmetry. In this
section, we briefly review our chaotic inflation model \cite{KYY}.

For the inflaton chiral supermultiplet $\Phi(x,\theta)$, we assume
that the K\"ahler potential $K(\Phi,\Phi^{\ast})$ is invariant under
the shift of $\Phi$,\footnote{The inflaton $\Phi$ may be one of
modulus fields in string theories. We hope that the explicit breaking
of the shift symmetry introduced below will be understood by yet
unknown dynamics of string theories.}
\bea
  \Phi \rightarrow \Phi + i~C M_{G},
  \label{eq:shift}
\eea
where $C$ is a dimensionless real parameter. Hereafter, we set $M_{G}$
to be unity. Thus, the K\"ahler potential is a function of $\Phi +
\Phi^{\ast}$, i.e. $K(\Phi,\Phi^{\ast}) = K(\Phi + \Phi^{\ast})$. It
is now clear that the supergravity effect $e^{K(\Phi + \Phi^{\ast})}$
discussed above does not prevent the imaginary part of the scalar
components of $\Phi$ from having a value larger than the gravitational
scale. So, we identify it with the inflaton field $\varphi$ [see Eq.
(\ref{eq:field})]. As long as the shift symmetry is exact, the
inflaton $\varphi$ never has a potential and hence it never causes
inflation. Therefore, we need some breaking term in the
superpotential. Here, we discuss the form of the superpotential.
First of all, we assume that in addition to the shift symmetry, the
superpotential is invariant under the $U(1)_{\rm R}$ symmetry, which
prohibits a constant term in the superpotential.  Then, the above
K\"ahler potential is invariant only if the R charge of $\Phi$ is
zero. Therefore, the superpotential comprised of only the $\Phi$ field
is not invariant under the U$(1)_{\rm R}$ symmetry, which compels us
to introduce another supermultiplet $X(x,\theta)$ with its R-charge
equal to two.

We now introduce a suprion field $\Xi$ describing the breaking of the
shift symmetry, and extend the shift symmetry including the suprion
field $\Xi$ as follows,\footnote{If $\Xi$ transforms as $\Xi
\rightarrow \frac{\Phi^{n}}{(\Phi + iC)^{n}} \Xi$ $(n \ge 2)$, we
have $W = X \Xi\Phi^{n}$, which may cause $\varphi^{2n}$ chaotic
inflations.}
\bea
  \Phi &\rightarrow& \Phi + i~C \non, \\
  \Xi  &\rightarrow& \frac{\Phi}{\Phi + i~C} \Xi.
  \label{eq:shift2}
\eea
That is, the combination $\Xi\Phi$ is invariant under the shift
symmetry. Then, the general superpotential invariant under the shift
and U$(1)_{\rm R}$ symmetries is given by
\bea
  W = X \lhk \Xi\Phi + \alpha_{3}(\Xi\Phi)^{3} + \cdots \rhk
     + \delta_{1} X \lhk 1 + \alpha_{2}(\Xi\Phi)^{2} + \cdots \rhk ,
  \label{eq:superpotential2}
\eea
where we have assumed the R charge of $\Xi$ vanish. The shift symmetry
is softly broken by inserting the vacuum value $\la \Xi \ra = m$. The
mass parameter $m$ is fixed at a value much smaller than unity
representing the magnitude of breaking of the shift symmetry
(\ref{eq:shift2}). We see that higher order terms with $\alpha_{i}$ of
the order of unity become irrelevant for the dynamics of the chaotic
inflation. Thus, we neglect them in the following discussion
\footnote{Among all complex constants, only a constant becomes real by
use of the phase rotation of the $X$ field. Below we set $m$ to be
real.} unless explicitly mentioned. We should note that the complex
constant $\delta_{1}$ is also of the order of unity in general. But,
as shown later, the absolute magnitude of $\delta_{1}$ must be at most
of the order of $m$, which is much smaller than unity.  Therefore, we
introduce the $Z_{2}$ symmetry, under which both the $\Phi$ and $X$
fields are odd. Then, the smallness of the constant $\delta_{1}$ is
associated with the small breaking of the $Z_{2}$ symmetry.  That is,
we introduce a suprion field $\Pi$ with odd charge under the $Z_{2}$
symmetry. The vacuum value $\la \Pi \ra = \delta_{1}$ breaks the
$Z_{2}$ symmetry.  Though the above superpotential is not invariant
under the shift and the $Z_{2}$ symmetries, the model is completely
natural in the 't Hooft's sense \cite{tHooft} because we have enhanced
symmetries in the limit $m$ and $\delta_{1} \rightarrow 0$. We use, in
the following analysis, the superpotential,
\beq
  W \simeq mX\Phi + \delta_{1}X.
\eeq

The K\"ahler potential invariant under the shift and U$(1)_{\rm R}$
symmetries is give by
\beq
  K = \delta_{2}(\Phi + \Phi^{\ast}) 
     + \frac12~(\Phi + \Phi^{\ast})^{2} 
     + XX^{\ast}  + \cdots.
  \label{eq:kahler}
\eeq
Here $\delta_{2} \sim |\delta_{1}|$ is a real constant representing
the breaking effect of the $Z_{2}$ symmetry. The terms
$\delta_{3}m_{3}\Phi + \delta_{3}^{\ast}m_{3}^{\ast}\Phi^{\ast}$ and
$(m_{4}\Phi)^{2} + (m_{4}^{\ast}\Phi^{\ast})^{2}$ may appear, where
$\delta_{3}$ and $m_{4}$ are complex constants representing the
breaking of the $Z_{2}$ and the shift symmetries($|\delta_{3}| \sim
|\delta_{1}|$ and $|m_{3}| \sim |m_{4}| \sim m$). But, these terms are
extremely small so we have omitted them in the K\"ahler potential
(\ref{eq:kahler}). We have also omitted a constant term because it
only changes the overall factor of the potential, whose effect can be
renormalized into the constant $m$ and $\delta_{1}$. Here and
hereafter, we use the same characters for scalar with those for
corresponding supermultiplets.

%%%%%%%%%%%%%%%%%%%%%%%%%%%%%%%%%%%%%%%%%%%%%%%%%%%%%%%%%%%%%%%%%%%%%%
\section{Dynamics of chaotic inflation}
%%%%%%%%%%%%%%%%%%%%%%%%%%%%%%%%%%%%%%%%%%%%%%%%%%%%%%%%%%%%%%%%%%%%%%

\label{sec:dynamics}

The Lagrangian density $\CL(\Phi,X)$ neglecting the higher order terms
is given by
\beq
  \CL(\Phi,X) = \partial_{\mu}\Phi\partial^{\mu}\Phi^{\ast} 
  + \partial_{\mu}X\partial^{\mu}X^{\ast}
         -V(\Phi,X),
  \label{eq:Lagrangian}
\eeq
with the potential $V(\Phi,X)$,
\beq
  V(\Phi,X) = m^{2} e^{K} 
      \lkk 
      |X|^{2} 
      |1 + (\delta_{2} + \Phi + \Phi^{\ast})(\Phi + \delta_{1}') |^{2}
      + |\Phi + \delta_{1}'|^{2} (1 - |X|^{2} + |X|^{4}) 
      \rkk ,
      \label{eq:potential}
\eeq
with $\delta_{1}' \equiv \delta_{1}/m$. Now, we decompose the complex
scalar field $\Phi$ into two real scalar fields as
\beq
  \Phi = \frac{1}{\sqrt{2}} (\eta + i \varphi),
  \label{eq:field}
\eeq
where we identify $\varphi$ with the inflaton. Then, the Lagrangian
density $\CL(\eta,\varphi,X)$ is given by
\beq
  \CL(\eta,\varphi,X) = \frac{1}{2}~\del_{\mu}\eta\del^{\mu}\eta 
              + \frac{1}{2}~\del_{\mu}\varphi\del^{\mu}\varphi 
              + \del_{\mu}X\del^{\mu}X^{*}
              -V(\eta,\varphi,X),
  \label{eq:Lagrangian2}
\eeq
with the potential $V(\eta,\varphi,X)$,
\bea
  \lefteqn{V(\eta,\varphi,X) = m^{2}e^{-\frac{\delta_{2}^{2}}{2}}
           \exp \lhk \lmk \eta + \frac{\delta_{2}}{\sqrt{2}} \rmk^{2} 
                     + |X|^{2}
                \rhk} \non \\ 
     &&    \times \lkk
            |X|^{2} 
             \lhk 1 + 2 \lmk \eta + \frac{\delta_{2}}{\sqrt{2}}\rmk
                              (\eta + \delta_{\rm R})
                    + \lmk \eta + \frac{\delta_{2}}{\sqrt{2}} \rmk^{2}
                    \lkk (\eta + \delta_{\rm R})^{2}
                      + (\varphi + \delta_{\rm I})^{2}  
                    \rkk
             \rhk \right.\non \\
     &&    \left.
           + \frac12~\lhk
                        (\eta + \delta_{\rm R})^{2}
                        + (\varphi + \delta_{\rm I})^{2}  
             \rhk (1 - |X|^{2} + |X|^{4}) 
                  \rkk.
  \label{eq:potential2}
\eea
Here, the complex constant $\delta_{1}'$ is decomposed into a real and
an imaginary part,
\beq
  \delta_{1}' = \frac{1}
     {\sqrt{2}}(\delta_{\rm R} + i \delta_{\rm I}).
\eeq

Note that $\eta$ and $|X|$ should be taken as $|\eta|, |X| \lesssim
{\cal O}(1)$ for $\delta_{2} \ll 1$ because of the presence of $e^{K}$
factor. On the other hand, $\varphi$ can take a value much larger than
${\cal O}(1)$ since $e^{K}$ does not contain $\varphi$. For the case
$\eta, |X| \ll {\cal O}(1)$, which is valid during the inflation as
shown later, the potential can be approximated as
\beq
  V(\eta,\varphi,X) \simeq 
                 \frac{1}{2}~m^{2} \widetilde{\varphi}^{2}
                 + m^{2} |X|^2,
\eeq
where $\widetilde\varphi \equiv \varphi + \delta_{\rm I}$ and we have
taken $m e^{-\delta_{2}^{2}/4} \simeq m$ since $\delta_{2} \ll 1$.

Thus, the term proportional to $\widetilde\varphi^{2}$ becomes
dominant and the chaotic inflation takes place if the initial value
$\widetilde\varphi \gg 1$. The potential minimum for $\eta$ during the
inflation, $\eta_{m}$, is given by the minimum of the K\"ahler
potential, which yield $\eta_{m} \simeq -
\frac{\delta_{2}}{\sqrt{2}}$.  Then, during the chaotic inflation, the
effective mass squared of $\eta$, $m_{\eta}^{2}$, becomes
\beq
  m_{\eta}^{2} \simeq m^{2} \widetilde\varphi^{2} \simeq 6 H^{2},
  \label{eq:etamass}
\eeq
where $H[\simeq \frac{1}{\sqrt{6}}m\widetilde\varphi]$ is the Hubble
parameter. Because $m_{\eta}^{2}$ is much larger than $\frac94 H^{2}$,
the field $\eta$ rapidly oscillates around the minimum $\eta_{m}$ with
its amplitude damped in proportion to $a^{-3/2}$, where $a$ is the
scale factor. Thus, the field $\eta$ settles down to the minimum
$\eta_{m}$ very quickly.

On the other hand, the effective mass of $X$, $m_{X}$, is $m$, which
is smaller than the Hubble scale so that it does not oscillate but
only slow rolls.\footnote{If we take the higher order term $\xi
|X|^{4}$ with $\xi < -9/8$ in the K\"ahler potential, the effective
mass squared of $X$ becomes larger than $9H^{2}/4$ so that $X$ rapidly
oscillates around the origin and its amplitude goes to zero.} Using
the slow-roll approximation, the classical equations of motion for
both $\widetilde\varphi$ and $X$ fields are given by
\bea
    3H \dot{\widetilde\varphi} & \simeq & - m^2 \widetilde\varphi, 
                              \label{eq:varphieq} \\
    3H \dot{X} & \simeq & - m^2 X, \label{eq:Xeq}
\eea
where the overdot represents the time derivative. Also, here and
hereafter, we assume that $X$ is real and positive. Then, we obtain
the relation between $\widetilde\varphi$ and $X$ fields,
\beq
  \lmk \frac{X}{X(0)} \rmk \simeq  
  \lmk \frac{\widetilde\varphi}{\widetilde\varphi(0)} \rmk,
  \label{eq:Xvarphi}
\eeq
where $\widetilde\varphi(0)$ and $X(0)$ are the initial values of
$\widetilde\varphi$ and $X$ fields. But, one should note that this
relation actually holds if and only if quantum fluctuations are
unimportant for both $\widetilde\varphi$ and $X$ fields. Therefore, we
need to clarify when the classical description is feasible. For this
purpose, we first compare quantum fluctuations with classical changes
for the field $\widetilde\varphi$.  During one expansion time, by use
of Eqs. (\ref{eq:etamass}) and (\ref{eq:varphieq}), the classical
change $\delta\widetilde\varphi_{c}$ becomes
\beq
  \delta\widetilde\varphi_{c} \simeq |\dot{\widetilde\varphi}| H^{-1} 
    \simeq \frac{2}{\widetilde\varphi}.
\eeq 
On the other hand, the amplitude of quantum fluctuations
$\delta\widetilde\varphi_{q} \simeq H / (2\pi)$. Thus, the above
classical equation of motion for $\widetilde\varphi$ is valid only if
$\widetilde\varphi \ll \widetilde\varphi_{i} \equiv
\sqrt{4\pi\sqrt{6}/m}$. Otherwise, the universe is in a
self-reproduction stage of eternal inflation \cite{eternal,sr} and the
current horizon scale is contained in a domain where
$\widetilde\varphi$ got smaller than $\widetilde\varphi_{i}$ and the
classical description of $\widetilde\varphi$ with the above classical
equation of motion became feasible. Therefore, we consider only the
region $\widetilde\varphi \ll m^{-1/2}$.\footnote{In this region we
may safely neglect the higher order terms of $\Xi\Phi$ in
Eq. (\ref{eq:superpotential2}).}

Next, in order to estimate the amplitude of quantum fluctuations of
$X$, we use the Fokker-Planck equation for the statistical
distribution function of $X$, $P[X,t]$,
\beq
  \frac{\del}{\del t}P[X,t]
    = \frac{1}{3H(t)}\frac{\del}{\del X} \lmk m^2XP[X,t] \rmk
      + \frac{H^3(t)}{8\pi^2}\frac{\del^2}{\del X^2}P[X,t], 
    \label{eq:FPeq}
\eeq
which is obtained through the Langevin equation based on Eq.
(\ref{eq:Xeq}) with use of the stochastic inflation method of
Starobinsky \cite{stochastic}. Then, the time evolution of the root
mean square (RMS) of fluctuations of $X$ is given by
\beq
  \frac{d}{dt} \la \lmk\Delta X\rmk^2 \ra = 
     - \frac{2 m^2}{3 H} \la \lmk\Delta X\rmk^2 \ra
      + \frac{H^{3}}{4\pi^{2}}.
  \label{eq:evXfluc}
\eeq
Taking $\widetilde\varphi$ as a time variable in Eq.
(\ref{eq:evXfluc}) by virtue of Eq. (\ref{eq:varphieq}), we find that
the RMS fluctuations of $X$ in an initially homogeneous domain at
$\widetilde\varphi=\widetilde\varphi_{i}$ are given by
\beq
   \la \lmk\Delta X\rmk^2 \ra = \frac{m^{2}}{96\pi^{2}}
     \lmk \widetilde\varphi_{i}^2\widetilde\varphi^2 -\widetilde\varphi^4 \rmk,  
      \label{eq:Xfluc}
\eeq
at the epoch $\widetilde\varphi$. Taking $\widetilde\varphi_{i} \simeq
\sqrt{4\pi\sqrt{6}/m}$, $\la \lmk\Delta X\rmk^2 \ra$ asymptotically
approaches
\beq
   \la \lmk\Delta X\rmk^2 \ra \simeq
     \frac{\sqrt{6}m}{24\pi}\widetilde\varphi^2.
      \label{eq:Xyuragi}
\eeq
On the other hand, from Eq. (\ref{eq:Xvarphi}), the classical value of
$X$, $X_{c}$ is at most $\widetilde\varphi/\widetilde\varphi_{i} \simeq
\sqrt{m}\widetilde\varphi/\sqrt{4\pi\sqrt{6}}$. Thus, during the chaotic
inflation,
\beq
  \sqrt{\la \lmk\Delta X\rmk^2 \ra} \lesssim X_{c} \sim \sqrt{m}\widetilde\varphi 
        \ll 1 \ll \widetilde\varphi = \varphi + \delta_{\rm I},
\eeq 
because $m \ll 1$ as shown later. Thus, for $X$, quantum fluctuations
are smaller than the classical value, and moreover our approximation
that both $\eta$ and $X$ are much smaller than unity is consistent
throughout the chaotic inflation.
 
Let us investigate the minimum of the potential after the chaotic
inflation. Since $X \sim \sqrt{m}(\varphi+\delta_{\rm I}) \ll 1$ as
shown above, the potential can be rewritten as\footnote{In fact, after
the inflation, the $X$ field also decays into standard particles so
that the amplitude of $X$ rapidly goes to zero. Hence, we can safely
set $X$ to be zero.}
\beq
  V(\eta,\varphi,X=0) = \frac12~m^{2} 
      \exp \lmk \sqrt{2} \delta_{2}\eta + \eta^{2} \rmk
       \lkk (\eta + \delta_{\rm R})^{2}
           + (\varphi + \delta_{\rm I})^{2} 
       \rkk.
\eeq
The extreme of the potential is obtained by the conditions
$\del V/\del \varphi = \del V/\del \eta = 0$, 
\bea
  \frac{\del V}{\del\varphi} &=& m^{2} 
      \exp \lmk \sqrt{2} \delta_{2}\eta^{2}+ \eta^{2} \rmk
       (\varphi + \delta_{\rm I})
 = 0, \non \\
 \frac{\del V}{\del\eta} &=& m^{2} 
      \exp \lmk \sqrt{2} \delta_{2}\eta^{2}+ \eta^{2} \rmk 
       \lhk \lmk \eta + \frac{\delta_{2}}{\sqrt{2}} \rmk
         \lkk (\eta + \delta_{\rm R})^{2}
           + (\varphi + \delta_{\rm I})^{2} 
         \rkk + (\eta + \delta_{\rm R})
       \rhk = 0,
\eea
which yields $\varphi = - \delta_{\rm I}$ and 
\beq
 (\eta + \delta_{\rm R})
    \lhk 2\eta^{2}+(2\delta_{\rm R}+\sqrt{2}\delta_{2})\eta
         +\sqrt{2}\delta_{\rm R}\delta_{2}+1 \rhk=0.
\eeq
Thus, for $|\sqrt{2}\delta_{\rm R} - \delta_{2}| \le 4$, $\eta = -
\delta_{\rm R}$ is only a minimum of the potential. Otherwise, there
is another local minimum near the minimum during the inflation, which
generally prevents the inflation from ending. Hence, the condition
that $\delta_{2} \sim |\delta_{1}| \lesssim m \sim 10^{-5}$ must be
satisfied for a successful inflation.

Now that preparations are complete, the density fluctuations produced
by this chaotic inflation is estimated as \cite{PS}
\beq
  \frac{\delta \rho}{\rho}\simeq \frac{1}{5\sqrt{3}\pi}
     \frac{m}{2\sqrt{2}} \lkk (\varphi+\delta_{\rm I})^2 + X^2 \rkk.
\eeq
Since $X \ll \varphi+\delta_{\rm I}$ as shown above, the amplitude of
the density fluctuations is actually determined only by the $\varphi$
field. Then, the normalization at the COBE scale [$\delta\rho/\rho
\simeq 2\times 10^{-5}$ for $(\varphi+\delta_{\rm I})_{\rm COBE}
\simeq 14$ \cite{COBE}] gives \footnote{The spectral index $n_{s}
\simeq 0.96$ for $(\varphi+\delta_{\rm I})_{\rm COBE} \simeq 14$.}
\beq
   m \simeq 10^{13}~{\rm GeV} \simeq 10^{-5}.
\eeq

After the inflation ends, the inflaton field $\varphi$ begins to
oscillate and its successive decays cause reheating of the universe.
The reheating may take place by introducing the following
superpotential:
\beq
    W = \delta_{4} X H_{u} H_{d},
    \label{eq:hhbarsuper}
\eeq
where $\delta_{4} = g \la \Pi \ra$ is a constant associated with the
breaking of the $Z_{2}$ symmetry. For $g = \CO(1)$, $\delta_{4} \sim
|\delta_{1}| \lesssim m \sim 10^{-5}$ as shown above. $H_{u}$ and
$H_{d}$ are a pair of Higgs doublets. Taking the R-charge and the
$Z_{2}$ charge of $H_{u}H_{d}$ to be zero and positive, the above
superpotential is invariant under the $U(1)_{\rm R}$ symmetry.

Then, we have a coupling of the inflaton $\varphi$ to the Higgs boson
doublets as
\beq
    L \simeq  \delta_{4} m \widetilde\varphi H_{u} H_{d},
    \label{eq:hhbarint}
\eeq
which gives the reheating temperature $T_{\rm RH}$\footnote{Field X
decays into the Higgsinos $\widetilde{H_{u}}$ and $\widetilde{H_{d}}$
through the Yukawa interaction in Eq. (\ref{eq:hhbarsuper}) with the
similar decay rate. Thereafter, field X rapidly goes to zero so that a
pair of Higgs doublets do not acquire additional masses.}
\beq
    T_{\rm RH} \lesssim 10^{9}~{\rm GeV} 
    \lmk \frac{\delta_{4}}{10^{-5}} \rmk
    \lmk \frac{m}{10^{13}~{\rm GeV}} \rmk^{1/2}.
\eeq
Since $\delta_{4} \lesssim m \sim 10^{-5}$, the reheating temperature
$T_{\rm RH}$ becomes less than $10^{9}$ GeV. Such a reheating
temperature is low enough to avoid the gravitino problem. Recently,
nonthermal production at the preheating stage was found to be
important in some inflation models \cite{Kallosh}. For the present
model, as shown by Kallosh {\it et al.} \cite{Kallosh}, nonthermal
production of gravitinos at the preheating phase is roughly estimated
as
\bea
  \lmk \frac{n_{3/2}}{s} \rmk_{\rm nonTH} 
   \sim \, \frac{m^{3}}{m^{2}/T_{R}} \,
    \lesssim 10^{-14} \lmk \frac{T_R}{10^{9}~{\rm GeV}} \rmk
                      \lmk \frac{m}{10^{13}~{\rm GeV}} \rmk,
\eea
where $n_{3/2}$ and $s$ are the number density of gravitinos and
entropy density. This is much less than the thermal production given
by $(n_{3/2}/s)_{\rm TH} \sim 10^{-12}(T_R/10^{9}~{\rm GeV})$ and
hence we can neglect the nonthermal production of gravitinos.

%%%%%%%%%%%%%%%%%%%%%%%%%%%%%%%%%%%%%%%%%%%%%%%%%%%%%%%%%%%%%%%%%%%%%%
\section{Leptogenesis via the inflaton decay in chaotic inflation}
%%%%%%%%%%%%%%%%%%%%%%%%%%%%%%%%%%%%%%%%%%%%%%%%%%%%%%%%%%%%%%%%%%%%%%

\label{sec:lepto}

In this section, we discuss the leptogenesis scenario via the inflaton
decay in the above chaotic inflation model. Many leptogenesis
scenarios have been proposed, so far, depending on the production
mechanisms of heavy Majorana neutrinos $N_{i}$
\cite{FY,LGthermal,LGinfdec,LGinfdec2,LGinfdec3,LGosc}. One of the
most attractive scenarios is the thermal production of heavy Majorana
neutrinos $N_{i}$ ($i$ = 1 - 3:the family index) during the reheating
stage after inflation. Detailed analyses \cite{LGthermal}, however,
show that enough lepton asymmetry is produced to explain the observed
baryon number density only if the reheating temperature is as high as
$10^{10}$~GeV.\footnote{In our model, when $\delta_{4} = g \la \Pi \ra
\sim 10^{-4}$ with $g = \CO(10)$, the reheating temperature $T_{\rm
RH}$ becomes as high as $10^{10}$~GeV so that the thermal production
of heavy Majorana neutrinos $N_{i}$ leads to enough lepton asymmetry
to explain the observed baryon number density.} Such a high reheating
temperature may cause the gravitino problem unless the gravitino mass
is very light ($\lesssim 1$~KeV) \cite{light} or very heavy ($\gtrsim
3$~TeV) \cite{heavy}.\footnote{Another solution where the gravitino is
the lightest supersymmetric particle of masses from 10 to 100 GeV is
proposed \cite{BBP}.} Another interesting scenario\footnote{Giudice
{\it et al.} discussed the production of heavy Majorana neutrinos
during preheating and the successive leptogenesis \cite{LGinfdec2}.
But, in our model, as given later in Eq. (\ref{eq:Nsuperpotential}),
the Yukawa coupling of the inflaton with heavy Majorana neutrinos is
so small that sufficient lepton asymmetry cannot be produced to
explain the observed baryon number density.} is that heavy Majorana
neutrinos $N_{i}$ are produced nonthermally via the decay of the
inflaton \cite{LGinfdec2,LGinfdec3}. We consider, here, a leptogenesis
scenario via the inflaton decay in the above mentioned chaotic
inflation model.\footnote{In Ref.  \cite{BB}, the direct baryogenesis
scenario via inflaton decay is discussed in the context of the chaotic
inflationary model in SU$(1,1)$ $N=1$ supergravity \cite{GL}.}

For our purpose, we extend the $Z_{2}$ symmetry into a $Z_{4}$
symmetry.  The charges of the $Z_{4}$ symmetry for various
supermultiplets are given in Table \ref{tab:charges}. Then, we
introduce the following superpotential invariant under the $U(1)_{R}$
and the $Z_{4}$ symmetries:
\beq
 W = \lambda_{i} m \Phi N_{i} N_{i} + \gamma_{i} \Pi N_{i} N_{i},
 \label{eq:Nsuperpotential}
\eeq
where $\lambda_{i}$ and $\gamma_{i}$ are constants and $\Pi$ is the
suprion field introduced before, whose vacuum value $\la \Pi \ra$
leads to the breaking of the $Z_{4}$ symmetry\footnote{This $Z_{4}$
symmetry is broken down to another $Z_{2}$ symmetry by $\la \Pi \ra
\ne 0$, where this $Z_{2}$ symmetry is nothing but the so-called
matter parity.} and must be less than $m \sim 10^{-5}$. Here, we set
$\la \Pi \ra \sim m \sim 10^{-5}$. The Majorana masses of right-handed
neutrinos $M_{i}$ is given by $M_{i} = \gamma_{i} \la \Pi \ra$.  For
$\gamma_{3} = \CO(1)$, $M_{3} \sim 10^{-5} \sim 10^{13}$~GeV. The
inflaton $\varphi$ and the orthogonal field $\eta$ can decay into
right handed scalar neutrinos $N_{i}$ through the above Yukawa
interactions if $M_{i} < m /2$. Both decay rates are similar and given
by
\beq
  \Gamma_{\varphi} \simeq \Gamma_{\eta} \simeq 
        \lambda^{2} \frac{m^{3}}{32\pi} 
         \sim 10 \lambda^{2}~{\rm GeV},
  \label{eq:varphidecay}
\eeq
with $\lambda^{2} \equiv \Sigma \lambda_{i}^{2}$ and $i$ runs for
$M_{i} \ll m$.\footnote{The field $X$ decays into $\widetilde{N}_{i}$
(sneutrinos) through the cross term of the superpotential with the
similar decay rate to Eq. (\ref{eq:varphidecay}). $\varphi$ and $\eta$
also have the decay channel into $\widetilde{N}_{i}$ but their decay
rates are much smaller than $\Gamma_{\varphi}$ and $\Gamma_{\eta}$.}
Then, the reheating temperature $T_{\rm RH}$ is given by
\beq
  T_{\rm RH} \sim 10^{9} \lambda~{\rm GeV}.
\eeq
For $\lambda < g$, the decay into $H_{u}H_{d}$ [see Eq.
(\ref{eq:hhbarint})] becomes the dominant decay mode of the inflaton
so that the reheating temperature becomes $T_{\rm RH} \sim 10^{9}
g$~GeV and the branching ratio of the decay into right-handed
neutrinos becomes $\CO(\lambda^{2}/g^{2})$ because $\delta_{4} = g \la
\Pi \ra \sim gm \sim 10^{-5}g$.

The produced $N_{i}$ decay into leptons $l_{j}$ and Higgs doublets
$H_{u}$ through the following Yukawa interactions of Higgs
supermultiplets, which is invariant under the U$(1)_{R}$ and the
$Z_{4}$ symmetries;
\beq
  W = (h_{\nu})_{ij}N_{i}l_{j}H_{u}.
\eeq
Here we have taken a basis where the mass matrix for $N_{i}$ is
diagonal, and have assumed that quarks and leptons can be classified
into the SU$(5)$ multiplets, ${\bf 10} = (q,u^{c},e^{c})$, ${\bf 5} =
(d^{c},l)$, and ${\bf 1} = (N)$. We also assume $|(h_{\nu})_{i3}| >
|(h_{\nu})_{i2}| \gg |(h_{\nu})_{i1}|$ (i = 1, 2, 3). We consider only
the decay of $N_{1}$ assuming that the mass $M_{1}$ is much smaller
than the others, $M_{2}$ and $M_{3}$.  The decay of $N_{1}$ has two
decay channels,
\bea
  N_{1} &\rightarrow& H_{u} + l, \\
        &\rightarrow& \overline{H_{u}} + \overline{l}.
\eea

These decay channels have different branching ratios if $CP$ symmetry
is violated. Interference between the tree-level and the one-loop
diagrams including vertex and self-energy corrections generates lepton
asymmetry \cite{FY,CRV,FPS,BP},
\bea
  \epsilon_1 
    &\equiv&
    \frac{ \Gamma (N_1 \rightarrow H_u + l )
         - \Gamma (N_1 \rightarrow \overline{H_u} + \overline{l} ) }
         { \Gamma_{N_1} }
    \non \\
    &=& - 
    \frac{3}{ 16 \pi \left( h_\nu h_\nu^{\dagger} \right)_{11} }
    \lkk 
        \mbox{Im} \left( h_\nu h_\nu^{\dagger} \right)_{13}^2 
        \frac{M_1}{M_3}
        +
        \mbox{Im} \left( h_\nu h_\nu^{\dagger} \right)_{12}^2 
        \frac{M_1}{M_2}
    \rkk.
    \label{eqn:cpasym}
\eea
By use of the above hierarchy of the Yukawa coupling constants, the
lepton asymmetry is dominated by the first term for
$\frac{m_{\nu_{3}}}{m_{\nu_{2}}} \gtrsim
\frac{M_{3}}{M_{2}}$\footnote{Even if the second term dominates, the
discussion also runs parallel.} and given by
\bea
  \epsilon_1 &\simeq& -
    \frac{3\delta_{\rm eff}}
         {16 \pi \lmk h_\nu h_\nu^{\dagger} \rmk_{11}}
    \left|\lmk h_\nu h_\nu^\dagger \rmk^2_{13} \right|
    \frac{M_1}{M_3} \non \\
            &\simeq& - 
    \frac{3\delta_{\rm eff}}
         {16 \pi}
    \left|\lmk h_\nu \rmk^2_{33} \right|
    \frac{M_1}{M_3} \non \\
            &\simeq& - 
    \frac{3\delta_{\rm eff}}
         {16 \pi}    
    \frac{m_{\nu_{3}}M_{1}}{\la H_{u} \ra^{2}} \non \\
            &\sim& - 
    10^{-5} \delta_{\rm eff} \lmk \frac{M_{1}}{10^{11}~\rm GeV} \rmk,
\eea
where $\delta_{\rm eff}$ is a parameter representing the magnitude of
the $CP$ violation, $m_{\nu_{3}}$ is estimated by the seesaw mechanism
\cite{seesaw} as
\bea
  m_{\nu_{3}} &\simeq& \frac{\left|\lmk h_\nu \rmk^2_{33}\right|
                           \la H_{u} \ra^{2}}{M_{3}} \non \\
              &\sim& 10^{-2}~{\rm eV} \lmk 
                       \frac{\left|\lmk h_\nu\rmk_{33}\right|}
                            {10^{-1}} \rmk
                             \lmk \frac{10^{13}~\rm GeV}{M_{3}} \rmk,
\eea
which is consistent with the mass suggested from the Super-kamiokande
experiments \cite{SK} for $|\lmk h_\nu\rmk_{33}| \sim 10^{-1}$ and
$M_{3} \sim 10^{13}$~GeV.

The total decay rate of $N_{1}$, $\Gamma_{N_1}$, is given by
\bea
  \Gamma_{N_1} &=& \Gamma (N_1 \rightarrow H_u + l )
         + \Gamma (N_1 \rightarrow \overline{H_u} + \overline{l} )
                   \non \\
               &\simeq& \frac{1}{8\pi} \Sigma |(h_{\nu})_{1i}|^{2}
                         M_{1} \non \\
               &\simeq& \frac{1}{8\pi} |(h_{\nu})_{13}|^{2} M_{1} 
                         \non \\
               &\sim& 10^{5}~{\rm GeV} 
                      \lmk \frac{|(h_{\nu})_{13}|}{10^{-2}} \rmk^{2}
                      \lmk \frac{M_{1}}{10^{11}~\rm GeV} \rmk.
\eea
Thus, for a wide range of parameters, the decay rate $\Gamma_{N_1}$ is
much larger than the decay rate of the inflaton $\Gamma_{\varphi}$ so
that the produced $N_{1}$ immediately decays into leptons and Higgs
supermultiplets.

Before estimating the lepton asymmetry produced in our model, let us
evaluate the lepton asymmetry needed to explain the observed baryon
number density. A part of produced lepton asymmetry, exactly speaking,
$B-L$ asymmetry is converted into baryon asymmetry through the
sphaleron processes, which can be estimated as \cite{KSHT}
\beq
  \frac{n_B}{s} \simeq - \frac{8}{23} \frac{n_L}{s},
\eeq
where we have assumed the standard model with two Higgs doublets and
three generations. In order to explain the observed baryon number
density,
\beq
  \frac{n_B}{s} \simeq (0.1 - 1) \times 10^{-10},
\eeq
we need the lepton asymmetry,
\beq
  \frac{n_L}{s} \simeq - (0.3 - 3) \times 10^{-10}.
\eeq

Now we estimate the lepton asymmetry produced through the inflaton
decay. For $M_{1} \gtrsim 10^{11} \lambda$~GeV, $M_{1}$ is one hundred
times larger than the reheating temperature $T_{\rm RH}$. In this
case, the produced $N_{1}$ is never in thermal equilibrium. Then, the
ratio of the lepton number to entropy density can be estimated as
\bea
  \frac{n_L}{s} 
          &\simeq&
       \frac32~\epsilon_1 B_r \frac{T_R}{m} 
    \non \\
    &\sim& 
    - 10^{-7} \delta_{\rm eff} B_r
    \lmk \frac{T_R}{10^{9}~\mbox{GeV}} \rmk
    \lmk \frac{M_1}{m} \rmk
    \non \\
    &\sim& 
    - 10^{-9} \delta_{\rm eff} B_r
    \lmk \frac{T_R}{10^{9}~\mbox{GeV}} \rmk
    \lmk \frac{M_1}{10^{11}~\mbox{GeV}} \rmk
    \lmk \frac{10^{13}~\mbox{GeV}}{m} \rmk,
\eea
where $B_r$ is the branching ratio of the inflaton decay into $N_1$.
For $M_{3} \sim M_{2} \sim m \sim 10^{13}$~GeV, the decay into $N_{3}$
and $N_{2}$ are prohibited kinematically or suppressed by the phase
space and hence $B_{r} = \CO(1)$ for $\lambda_{1} = \CO(1)$. In this
case, we obtain $T_{\rm RH} \sim 10^{9}$~GeV, which results in $n_L/s
\sim - 10^{-9} \delta_{\rm eff}$.  Thus, our model of leptogenesis
works well for $\gamma_{2} \simeq \gamma_{3} = \CO(1)$, $\delta_{\rm
eff} = \CO(1)$, and $\lambda_{1} = \CO(1)$ [see Eq.
(\ref{eq:Nsuperpotential})].

Finally, we make a comment on the Froggatt-Nielsen (FN) mechanism
\cite{FN} based on a spontaneously broken U$(1)_{F}$ family symmetry,
which gives a natural explanation for the observed mass hierarchy in
mass matrices of quarks and charged leptons. The U$(1)_{F}$ symmetry
is broken by a gauge singlet scalar field $\Delta$ with FN charge
$Q_{\Delta} = -1$, whose condensation $\la \Delta \ra$ gives rise to
the Yukawa coupling constants. That is, the Yukawa couplings of Higgs
supermultiplets are given through nonrenormalizable interactions with
$\Delta$,
\beq
  W = g_{ij}  \Delta^{Q_i + Q_j } \Psi_i \Psi_j H_{u(d)},
\eeq
where $Q_i$ are the FN charges of quark and lepton supermultiplets
$\Psi_i$, $g_{ij}$ are coupling constants of the order unity, and
$H_u$, $H_d$ are Higgs supermultiplets with FN charges zero. In
particular, $(h_{\nu})_{ij} = g_{ij} \la \Delta \ra^{Q_{N_{i}} +
Q_{l_{j}}}$.  Then, the observed mass hierarchy can be well explained
if we take $\epsilon \equiv \la \Delta \ra \simeq 1/17$ and the FN
charges of quark and lepton supermultiplets shown in Table
\ref{tab:charges2} \cite{MH}.

If the Froggatt-Nielsen mechanism is adopted, the above discussion on
the leptogenesis also holds except for three points. First of all, two
contributions to the lepton asymmetry in Eq. (\ref{eqn:cpasym}) become
comparable. Next, the coupling constants $\gamma_{i}$ in Eq.
(\ref{eq:Nsuperpotential}) becomes $\gamma_{3} \sim \gamma_{2} =
\CO(1)$ and $\gamma_{1} = \CO(10^{-2})$. Therefore, $M_{3}$ and $M_{2}
\sim 10^{13}$~GeV automatically become comparable with the mass of
inflaton $\varphi$ and other fields $\eta$ and $X$, i.e. $\sim m \sim
10^{13}$~GeV, so that the decay into $N_{3}$ and $N_{2}$ are
prohibited kinematically or suppressed by the phase space. Finally,
$\lambda_{1}$ in Eq. (\ref{eq:Nsuperpotential}) becomes $\lambda_{1} =
\CO(10^{-2})$ so that the reheating temperature $T_{\rm RH}$ becomes
$10^{7}$~GeV. In this case, unless $g < \CO(10^{-2})$, the decay mode
into the Higgs doublet in Eq. (\ref{eq:hhbarint}) must be forbidden
because otherwise the branching ratio becomes small as $B_{r} \sim
\lambda_{1}^{2}/g^{2}$ and the produced lepton asymmetry may be too
small. If, for example, we set the R-charge of $H_{u}H_{d}$ to be
nonzero,\footnote{In this case, other necessary Yukawa interactions
are all permitted, taking the R-charges of $H_{u}$, $H_{d}$, ${\bf
5}^{\ast}$, and ${\bf 10}$ to be $2a/5$, $3a/5$, $1-2a/5$, and
$1-a/5$, where the R-charge of $H_{u}H_{d}$ is $a$.} the
superpotential in Eq. (\ref{eq:hhbarsuper}) is prohibited. Then, the
ratio of lepton number density to entropy density can be estimated as
\bea
  \frac{n_L}{s} 
    \sim 
    - 10^{-11} 
    \lmk \frac{T_R}{10^{7}~\mbox{GeV}} \rmk
    \lmk \frac{M_1}{10^{11}~\mbox{GeV}} \rmk
    \lmk \frac{10^{13}~\mbox{GeV}}{m} \rmk,
\eea
which is marginally consistent with the baryon number density in the
present universe.

%%%%%%%%%%%%%%%%%%%%%%%%%%%%%%%%%%%%%%%%%%%%%%%%%%%%%%%%%%%%%%%%%%%%%%
\section{Discussion and conclusions}
%%%%%%%%%%%%%%%%%%%%%%%%%%%%%%%%%%%%%%%%%%%%%%%%%%%%%%%%%%%%%%%%%%%%%%

\label{sec:con}

In this paper we have comprehensively investigated a natural chaotic
inflation model with the shift symmetry in supergravity. In
particular, the forms of the K\"ahler potential and the superpotential
have been discussed. In order to suppress higher order terms of the
inflaton field in the superpotential, the shift symmetry is extended
into that including the suprion field $\Xi$ with the combination
$\Xi\Phi$ invariant. Also, the linear term of $X$ in the
superpotential is suppressed by introducing the $Z_{2}$ symmetry. We
have found that if the magnitude of the breaking of the $Z_{2}$
symmetry is equal or smaller than that of the shift symmetry, a
desired chaotic inflation can take place.

We have also discussed the leptogenesis via the inflaton decay in this
chaotic inflation model. The inflaton $\varphi$ can decay into
right-handed neutrinos through the Yukawa interactions suppressed by
the breaking of the shift symmetry, which leads to low reheating
temperature enough to avoid the overproduction of gravitinos.
Right-handed neutrinos acquire their masses associated with the
breaking of a $Z_{4}$ symmetry which is an extension of the $Z_{2}$
symmetry, whose magnitude is consistent with the result from the
Super-kamiokande experiment.  Then, we have found that for a wide range
of parameters, the lepton asymmetry enough to explain the observed
baryon number density is produced. Also, when the Froggatt-Nielsen
mechanism is adopted as the mechanism to explain the hierarchy for the
masses of leptons and quarks, we have obtained the lepton asymmetry,
which is marginally consistent with the baryon number density in the
present universe.

%%%%%%%%%%%%%%%%%%%%%%%%%%%%%%%%%%%%%%%%%%%%%%%%%%%%%%%%%%%%%%%%%%%%%%
\subsection*{Acknowledgments}
%%%%%%%%%%%%%%%%%%%%%%%%%%%%%%%%%%%%%%%%%%%%%%%%%%%%%%%%%%%%%%%%%%%%%%

M.Y. is grateful to R. Brandenberger, K. Hamaguchi, A. A. Starobinsky,
E. D. Stewart, and J. Yokoyama for many useful discussions and
comments. M.K. and T.Y. are supported in part by the Grant-in-Aid,
Priority Area ``Supersymmetry and Unified Theory of Elementary
Particles''(\#707).  M.Y. is partially supported by the Japanese
Society for the Promotion of Science.

\begin{table}[t]
  \begin{center}
    \begin{tabular}{| c | c | c | c | c | c | c | c | c | c |}
                   & $\Phi$ & $X$ & $\Xi$ & $\Pi$ & $N$ & $H_{u}$
                   & $H_{d}$ & ${\bf 5^{\ast}}$ & ${\bf 10}$ \\
        \hline
        $Q_R$      & 0      & 2   & 0     & 0     & 1   & $0$
                   & $0$     & $1$              & $1$ \\ 
        \hline 
        $Z_{4}$    & $2$    & $2$ & $0$   & $2$   & $1$ & $2$
                   & $2$     & $1$              & $1$   
    \end{tabular}
    \caption{The charges of various supermultiplets of U$(1)_{\rm
    R} \times Z_{4}$. Here, R charge of $H_{u}H_{d}$ is assigned to
    be $0$. All supermultiplets of quarks and leptons have the $Z_{4}$
    charge $1$ and Higgs supermultiplets $H_{u}$ and $H_{d}$ carry the
    $Z_{4}$ charge $2$.}
    \label{tab:charges}
  \end{center}
\end{table}

\begin{table}[t]
  \begin{center}
    \begin{tabular}{| c | c c c | c c c  | c c c | }
        $\Psi_i$ & \multicolumn{3}{c|}{${\bf 5} = (d^{c},l)$}
                 & \multicolumn{3}{c|}{${\bf 10} = (q,u^{c},e^{c})$}
                 & \multicolumn{3}{c|}{${\bf 1} = (N)$} \\
                 & ${\bf 5}_3$  & ${\bf 5}_2$  & ${\bf 5}_1$ 
                 & ${\bf 10}_3$ & ${\bf 10}_2$ & ${\bf 10}_1$ 
                 & ${\bf 1}_3$  & ${\bf 1}_2$  & ${\bf 1}_1$ \\
        \hline
        $Q_i$    & 1     & 1     & 2
                 & 0     & 1     & 2
                 & 0     & 0     & 1 \\
    \end{tabular}
    \caption{The FN charges of quark and lepton supermultiplets
    assumed throughout this paper.}
    \label{tab:charges2}
  \end{center}
\end{table}

\end{document}